# Assessing a Hydrodynamic Description for Instabilities in Highly Dissipative, Freely Cooling Granular Gases

Peter P. Mitrano[a], Vicente Garzó[b], Andrew M. Hilger[a], Christopher J. Ewasko[a], Christine M. Hrenya[a],*

[a]*Department of Chemical & Biological Engineering, University of Colorado, Boulder, Colorado 80309, USA*

[b]*Departamento de Física, Universidad de Extremadura, E-06071 Badajoz, Spain*

An intriguing phenomenon displayed by granular flows and predicted by kinetic-theory-based models is the instability known as particle "clustering," which refers to the tendency of dissipative grains to form transient, loose regions of relatively high concentration. In this work, we assess a modified-Sonine approximation recently proposed [Garzó *et al.*, Physica A **376**, 94 (2007)] for a granular gas via an examination of system stability. In particular, we determine the critical length scale associated with the onset of two types of instabilities – vortices and clusters – via stability analyses of the Navier-Stokes-order hydrodynamic equations by using the expressions of the transport coefficients obtained from both the standard and the modified-Sonine approximations. We examine the impact of both Sonine approximations over a range of solids fraction $\phi$ <0.2 for small restitution coefficients $e$=0.25–0.4, where the standard and modified theories exhibit discrepancies. The theoretical predictions for the critical length scales are compared to molecular dynamics (MD) simulations, of which a small percentage were not considered due to inelastic collapse. Results show excellent quantitative agreement between MD and the modified-Sonine theory, while the standard theory loses accuracy for this highly dissipative parameter space. The modified theory also remedies a (high-dissipation) qualitative mismatch between the standard theory and MD for the instability that forms more readily. Furthermore, the evolution of cluster size is briefly examined via MD, indicating that domain-size clusters may remain stable or halve in size, depending on system parameters.

*Author to whom correspondence should be addressed

Instabilities, such as dynamic particle clusters, occur in both granular and gas-solid rapid flows [1-4]. In granular flows, such non-uniformities in concentration (bulk density) can be traced to the dissipative nature of collisions [1, 2, 5, 6], whereas in gas-solid flows, both mean drag and viscous damping can also lead to clustering instabilities [7-9]. Regardless of the source, clustering instabilities impact the performance of industrial operations, such as gas-solid fluidized beds and pneumatic conveyers [10]. Hydrodynamic descriptions derived from kinetic theory of rapid particulate flows predict inhomogeneities that are qualitatively similar to previous experiments and discrete-particle simulations [6, 11]. Here, we instead consider the quantitative ability of a modified kinetic theory applied to a simple granular system.

To date, much *qualitative* work has been done on instabilities in the homogeneous cooling of granular flows. The homogeneous cooling system (HCS) consists of dissipative particles in a periodic domain. The absence of external forces in the HCS gives rise to a homogeneous system when stable, though homogeneity may be lost due to the formation of velocity vortices (i.e., transversal- or shear-mode instabilities, as illustrated in Fig. 1b) or particle clusters (i.e., longitudinal- or heat-mode instabilities, as illustrated in Fig. 2b and 2c) [11-17]. More specifically, previous works have shown that the presence of such instabilities is dictated by the restitution coefficient $e$, the solids fraction $\phi$, and the ratio of the domain length to particle diameter $L/d$. Instabilities are more likely in larger domains, and, in general, velocity vortices manifest more readily than particle clusters [2, 17]. For a given $e$, $\phi$ pairing, a critical dimensionless system scale $L_{Vortex}/d$ demarcates (stable) homogeneous flow (Fig. 1a) from one with velocity-vortex instabilities (Fig. 1b), while a separate critical value $L_{Cluster}/d$ demarcates a (stable) homogeneous particle distribution (Fig. 2a) from one exhibiting the clustering instability (Fig. 2b & 2c) [11-19].

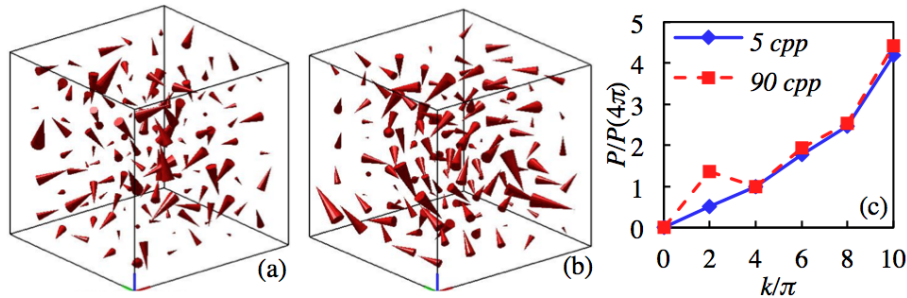

FIG. 1. (Color online) Spatially averaged velocity fields from MD simulations with $e=0.4$, $\phi=0.05$, and number of particles $N=2096$ at (a) 5 collisions per particle (*cpp*) and (b) 90 *cpp*, and (c) corresponding Fourier momentum spectra ($P$) normalized to the value at $k=4\pi$. Square cells with length $L/5$ were averaged for visualization in (a) and (b).

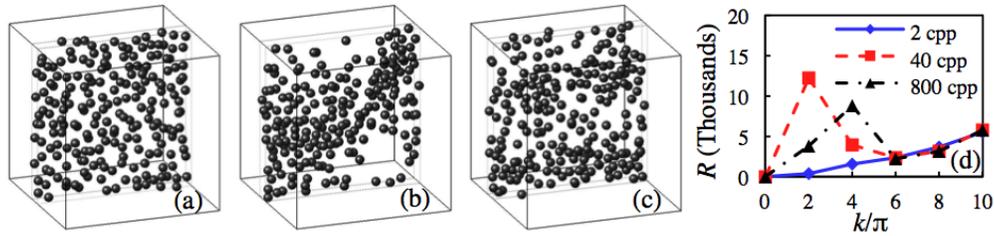

FIG. 2. (Color online) Snapshots of particle positions (showing only a slice of thickness $L/10$) extracted from three-dimensional MD simulations with $e=0.6$, $\phi=0.2$, $N=2000$ at (a) 2 *cpp*, (b) 40 *cpp*, (c) 800 *cpp*, and (d) corresponding Fourier mass spectra ($R$).

Recently, the first *quantitative* assessment [19] of predictions of the critical system size for velocity-vortex instabilities ($L_{Vortex}/d$) was made. The results show excellent agreement between molecular dynamics (MD) simulations and a linear stability analysis [17] of the hydrodynamic equations derived from the Enskog kinetic theory [20] over moderate ranges of dissipation ($0.6 \le e \le 0.9$) and solids fraction ($0.05 \le \phi \le 0.4$). It is worth noting that the difference between the previous treatment and the modified version lies in the modified kinetic theory used here, and how the integral equations defining the transport coefficients were evaluated. In particular, in order to obtain an analytical solution for integrals defining the transport coefficients, Ref. [20] approximated the first-order velocity distribution as the product of the Maxwell-Boltzmann distribution and a truncated Sonine polynomial; we will refer henceforth to this procedure as the *standard*-Sonine approximation. In spite of this simple approximation, such predictions compare well with Direct Simulation Monte Carlo (DSMC) results [21, 22], except for heat flux transport coefficients in highly dissipative systems ($e<0.6$) at the dilute limit. Motivated by this disagreement, a slight modification to the standard Sonine approximation has been recently proposed for monocomponent [23] and multicomponent [24] granular gases. The idea behind the *modified*-Sonine approximation is to assume that the isotropic part of the first-order distribution function is mainly governed by the HCS distribution rather than by the Maxwellian distribution. This modified-Sonine approach significantly improves the *e*-dependence of the heat flux transport coefficients and corrects the disagreement observed in dilute systems between theory and DSMC results. It is worth stressing that DSMC, a numerical solution of the Boltzmann or Enskog equation (from which the hydrodynamic equations and transport coefficients are derived analytically), is ideal for testing mathematical assumptions used in the derivation process, whereas MD, based on Newton's equations of motion, is entirely independent of the starting kinetic equation. MD thus serves as an ideal test bed for the kinetic equation itself along with assumptions used to derive the hydrodynamic equations.

In this work, we determine $L_{Vortex}/d$ and $L_{Cluster}/d$ via (the non-collapsed) MD simulations to quantitatively assess the ability of the standard [20] and modified-Sonine [23] approximations of the Enskog equation to predict vortex and cluster instabilities. We build on the earlier comparison [19], which was limited to velocity vortices, by considering a previously unexplored range of high dissipation ($e=0.25$-$0.4$) and dilute-to-moderate volume fractions ($\phi=0.05$-$0.15$), where the new modification [23] strongly impacts the theoretical transport coefficients. It is found that the modified theory performs considerably better than the standard one in predicting both the quantitative and qualitative nature of the instabilities in this region of high dissipation and moderate concentrations.

Monodisperse, frictionless, spherical particles with a constant, normal restitution coefficient $e$ are simulated via event-driven MD in a three-dimensional HCS. The solids fraction of the domain is given by

$$\phi = \frac{N\pi d^3}{6L^3},\quad(1)$$

where $N$ is the total number of particles, $d$ is the particle diameter, and $L$ is the length of the cubic domain. The assumption of instantaneous, binary collisions, which is shared by kinetic theory and MD, allows for straightforward comparison between theory and simulation. A more detailed description of hard-sphere MD is available in Refs. [25, 26] and the specific simulation method in Ref. [19].

Here, we ensure that the range of $e$ considered is low enough to explore the effects of the modifications [23] to the standard-Sonine approach [20], yet high enough that inelastic collapse (an unphysical artifact of the hard-sphere model [27-29]) does not interfere with our reported results. Specifically, we choose system parameters that allow at least 85% of simulation replicates to advance a minimum of 400 collisions per particle, or "*cpp*," without collapse; the results of collapsed systems are ignored.

Goldhirsch *et al.* [2] introduced a technique to characterize the intensity of instabilities in MD simulations of HCS based on a Fourier analysis. In particular, the Fourier transform of the momentum density $\hat{\underline{p}}$ and the mass density (concentration) $\hat{\rho}$ are given by [2]

$$\hat{\underline{p}}(\underline{k}) = \frac{m}{2\pi}\sum_{j=1}^{N}\underline{u}_j \exp(i\underline{k}\cdot\underline{x}_j),\quad(2)$$

$$\hat{\rho}(\underline{k}) = \frac{m}{2\pi}\sum_{j=1}^{N}\exp(i\underline{k}\cdot\underline{x}_j),\quad(3)$$

where $\underline{x}_j$ and $\underline{u}_j$ are the position and velocity of the $j^{th}$ particle, and $m$ is the particle mass. Periodic boundaries allow for the wavenumbers,

$$\underline{k} = \left(2a\pi/L_x,\ 2b\pi/L_y,\ 2c\pi/L_z\right),\quad(4)$$

where $a$, $b$, and $c$ are integers. Stability in the momentum and concentration fields can be deduced from the $\underline{k}$-dependence of the momentum spectrum $P$ and mass spectrum $R$, respectively,

$$P(\underline{k}) = \int_0^{2\pi} \int_0^{\pi} \int_0^{k+dk} |\hat{p}|^2 r^2 \sin\theta \, dr \, d\theta \, d\varphi, \tag{5}$$

$$R(\underline{k}) = \int_0^{2\pi} \int_0^{\pi} \int_0^{k+dk} |\hat{\rho}|^2 r^2 \sin\theta \, dr \, d\theta \, d\varphi, \tag{6}$$

where $P$ and $R$ are the norm squared of $\hat{p}$ and $\hat{\rho}$, respectively, integrated in spherical coordinates from a sphere with radius $k$ to $k+dk$ where $k = |\underline{k}|$. Here, we compute the corresponding Riemann sums for spherical shells (between each concentric, $\underline{k}$-space sphere) with an outer radius that is a factor of $2\pi$ greater than the inner radius, which is a 3-dimensional extension of the previous work [2, 30]. For a homogeneous system, $P$ and $R$ increase monotonically with $\underline{k}$ as the volume associated with each spherical shell increases. However, as noted by Goldhirsch *et al.* [2], unstable HCS displays a non-monotonic variation. This observation will be used as a basis for our stability criterion, as detailed below.

Figure 1 and 2 show snapshots of velocity and concentration fields, respectively, that are stable (Fig. 1a and 2a) and unstable (Fig. 1b, 2b, and 2c), along with Fourier spectra of momentum (Fig. 1c) and mass (Fig. 2d). Considering first the velocity-vortex instability (Fig. 1), note that a momentum mode ($P$) excitation appears at $k=2\pi$ (Fig. 1c) for the system exhibiting velocity vortices (Fig. 1b) but not for its stable counterpart (Fig. 1a). It follows that the presence of this excitation demarcates an unstable, vortex flow (Fig. 1b) from a stable, homogenous flow (Fig. 1a). This criterion agrees with a previous method [19] based on Haff's law. The Fourier spectra are measured throughout the simulation at intervals of 10 *cpp*, a value to which our results are insensitive. We begin averaging the percent difference (with respect to the $k=4\pi$ mode) between the excited-wavenumber ($k=2\pi$) mode and the following ($k=4\pi$) mode value once 3 consecutive measurements show $k=2\pi$ excitations in the momentum spectrum. If the averaged percent difference is positive, we deem the system unstable to vortices.

For each pairing of $e$, $\phi$ and $L/d$, 50 simulations are run with varied initial configurations. The "error" bars shown in Fig. 3 and 4 represent a bracketing of the critical value where the lower bound is the largest $L/d$ without any vortices present in the 50 simulations and the upper bound is the smallest $L/d$ with at least 1/50 simulations exhibiting vortices. The data points simply represent the average of the two bracketing $L/d$ values.

Detection of clustering instabilities (Fig. 2) follows a similar methodology as the velocity vortices (Fig. 1), though some interesting differences exist. While vortex structures were always observed to grow to the system size (i.e., no repeating vortex patterns), the same is not true for clusters. For relatively large $L/d$, vortices are sufficiently excited to shear the system-size concentration wave (i.e., a cluster and void region) into a wave with twice the frequency (i.e., two cluster and two void regions). This structural change gives to an excitation of the Fourier mass spectrum at $k=4\pi$ [2], a MD manifestation of the (non-linear) clustering theory derived by Soto *et al.* [14]. An example of this behavior is displayed in the particle position snapshot of Fig. 2c and the corresponding Fourier mass spectrum of Fig. 2d (i.e., 800 *cpp*). Note that this concentration wave with period=$L/2$ (Fig. 2d at 800 *cpp*) is preceded in time by the larger wave of period $L$ (Fig. 2c at 40 *cpp*). However, we do observe sustained clusters of period $L$ (as predicted by the linear theory [17] and analogous to Fig. 2b, but not evolving to higher-frequency clusters as in Fig. 2c) in simulations with domains slightly larger than $L_{Cluster}/d$. To the best of our knowledge, such clusters have not been displayed previously through MD simulations; previous work [2, 31] showed only higher-frequency, or smaller, clusters (see Fig. 2 of Ref. [2]) because large systems (i.e., $L/d >> L_{Cluster}/d$) were studied. In this work, care is taken to detect either type of sustained cluster (i.e., $k=2\pi$ and $k=4\pi$ excitations); with this exception, the criteria for $L_{Cluster}/d$ follow from the description of $L_{Vortex}/d$.

As mentioned previously, Garzó [17] used the Enskog kinetic equation to perform a linear stability analysis of the Navier-Stokes hydrodynamic equations. This stability analysis gave rise to critical system lengths associated with vortices and clusters,

$$L_{Vortex} = \frac{5\pi}{4}\sqrt{\pi}\chi\lambda_0\sqrt{\frac{\eta^*}{\zeta_0^*}}, \tag{7}$$

$$L_{Cluster} = \frac{5\pi}{4}\sqrt{\frac{5\pi}{2}}\chi\lambda_0\sqrt{\frac{\kappa^* C_p - \mu^*}{\zeta_0^*(2g - C_p)}}, \tag{8}$$

where $\chi$ is the pair correlation function at contact, $\lambda_0$ is the mean free path, $\eta^*$ is the dimensionless shear viscosity, $\zeta_0^*$ is the dimensionless zeroth-order cooling rate, $\kappa^*$ is the dimensionless thermal conductivity, $C_p$ and $g$ are functions of $\chi$ equal to unity in the dilute limit, and $\mu^*$ is the dimensionless heat-flux transport coefficient associated with gradients in concentration. Each of these quantities is defined in terms of $e$ and $\phi$; see Ref. [17] for detailed forms.

Equations (7) and (8) show that the critical lengths depend on the transport coefficients. Explicit forms for these coefficients have been obtained by means of the standard [20] and modified-Sonine [23] approximations. Both approaches differ significantly for small values of the restitution coefficient in the case of the heat flux transport coefficients. In this work, we use the standard [20] and modified [23] forms of the transport coefficients in Eq. (7) and (8) for the critical lengths and compare the predictions of $L_{Vortex}/d$ and $L_{Cluster}/d$ with MD data.

Unlike the previous evaluation of $L_{Vortex}/d$ using the standard theory at moderate dissipation and dilute-to-moderate concentrations [19], highly-dissipative systems are examined here. Focusing on this parameter space allows us to compare the standard and modified-Sonine approaches in the region where the discrepancy is expected to be most significant. Moreover, these results represent the first time that $L_{Cluster}/d$ has been obtained from MD simulations. This knowledge is important since predictions from the standard theory indicate that systems of moderate dissipation and volume fraction are more prone to the vortex instability, whereas highly dissipative, dilute systems are more prone to the clustering instability [17]. As detailed below, the MD and modified theory results indicate otherwise.

In Fig. 3, the critical system length scale for particle clustering ($L_{Cluster}/d$) is plotted for $e=0.25$ (Fig. 3a) and $e=0.4$ (Fig. 3b) over a range of $\phi=0.05\text{-}0.15$. Fig. 3a shows that the modified-Sonine approximation [23] improves the theoretical prediction for $L_{Cluster}/d$ relative to the standard prediction [17] in moderately dilute systems. This agreement provides evidence that the modified-Sonine approximation successfully accounts for non-Gaussian corrections to the particle velocity distribution, and that such corrections are important at high dissipation levels.

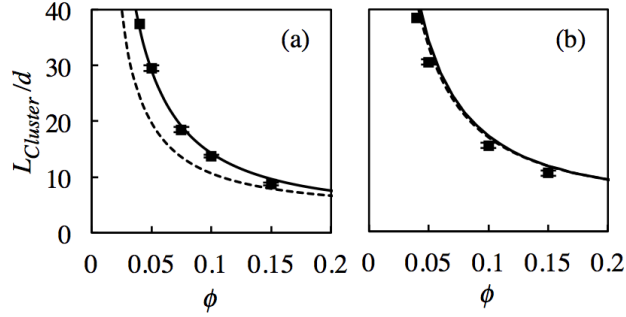

FIG. 3. Critical clustering length scale ($L_{Cluster}/d$) as a function of solids fraction for (a) $e=0.25$ and (b) $e=0.4$. The solid and dashed lines correspond to the modified and standard theories, respectively. The data points correspond to MD.

It is intriguing to note that Fig. 3b appears to suggest a negligible difference between the standard and modified theories for $e=0.4$ although theoretical transport coefficients (e.g., $\kappa^*$ or $\mu^*$) have been shown to differ for such dissipation [23]. The apparent contradiction can be explained by a closer examination of how the transport coefficients, namely $\kappa^*$ and $\mu^*$, impact $L_{Cluster}/d$. In particular, Garzó et al. [23] shows that both transport coefficients are overestimated by the standard theory [20], such that the deviation caused by the modified theory appears to "cancel out" in the computation of $L_{Cluster}/d$ via Eq. (8).

Figure 4 shows the $L_{Vortex}/d$ and $L_{Cluster}/d$ associated with the standard and modified theories, and MD for $e=0.25$. Note that the standard theory predicts $L_{Cluster}/d<L_{Vortex}/d$, while MD shows that vortices manifest at smaller $L/d$ than clusters. The modified-Sonine approach remedies this qualitative disparity in highly dissipative systems between MD and the standard theory; both MD and the modified theory suggest $L_{Vortex}/d<L_{Cluster}/d$.

Finally, the results presented here are robust. In particular, the quantitative agreement between MD and the modified theory shown in Fig. 3 and 4 is also representative of results obtained at $e=0.3$ and $\phi=0.05\text{–}0.15$, (not shown for the sake of brevity).

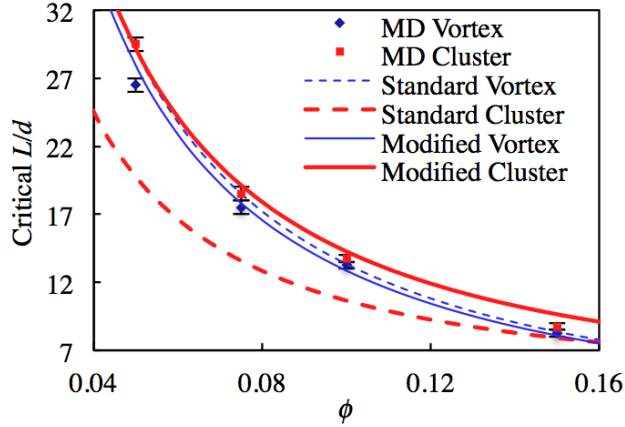

FIG. 4. (Color online) Critical length scale for vortex and cluster instabilities (i.e., $L_{Vortex}/d$ and $L_{Cluster}/d$, respectively) plotted as a function of solids fraction for $e=0.25$. The solid and dashed lines correspond to the modified and standard theories, respectively.

In summary, the present work reveals that a recent modification [23] to the standard-Sonine method [20] for obtaining the Navier-Stokes transport coefficients improves the quantitative predictive ability for highly dissipative ($e<0.4$) systems of dilute-to-moderate concentrations ($0.05 \leq \phi \leq 0.15$). Comparison with MD shows that the modified theory improves the quantitative prediction of $L_{Cluster}/d$ relative to the standard theory [20] and also correctly identifies the leading instability in a region where the standard theory does not.

The excellent quantitative agreement between theory and MD for the onset of vortices and clusters exemplifies the wide range of applicability of granular hydrodynamics derived from a modified kinetic theory and supports the claim that hydrodynamics (which is based on the separation between microscopic and macroscopic length and time scales) is not only limited to nearly elastic particles. Furthermore, the agreement with the Navier-Stokes theory (linear relations between fluxes and spatial gradients) is not entirely unexpected since the systems studied here are near the critical scale for instability such that large gradients have not yet developed. The present study, which is restricted to rigid-particle systems, also adds to a growing body of evidence [19, 32, 33] that the Enskog equation (which neglects velocity correlations between the particles that are about to collide) accurately describes moderately dilute and moderately dense systems. Finally, the results reported here are of practical interest since the high dissipation levels ($e=0.25$–0.4) examined in this work are also exhibited by wetted particles [34], which are critical in processes such as coagulation, spray coating, filtration, and pneumatic conveying.

P.P.M., A.M.H., C.J.E., and C.M.H. are grateful for the funding provided by the American Chemical Society under Grant No. PRF-50885-ND9 and the National Science Foundation REU site program under Grant No. NSF EEC-0851849. The research of V.G. has been supported by the Ministerio de Educación y Ciencia (Spain) through grant No. FIS2010-16587, partially financed by FEDER funds and by the Junta de Extremadura (Spain) through grant No. GR10158.